\documentstyle[psfig]{mn2e}

\title[X-ray variability of IRAS 13224--3809 from an {\em XMM-Newton} observation]
{The X-ray variability of the  Narrow-Line Seyfert 1 galaxy
IRAS 13224--3809 from an {\em XMM-Newton} observation }

\author[L. C. Gallo et al.]
{L. C. Gallo,$^1$ Th. Boller,$^1$ Y. Tanaka,$^{1,2}$ A. C. Fabian,$^3$ W. N. Brandt,$^4$ W. F. Welsh,$^5$
\newauthor N. Anabuki,$^2$ and Y. Haba$^2$\\ 
$^1$ Max-Planck-Institut f\"ur extraterrestrische Physik, Postfach 1312, 85741 Garching, Germany \\
$^2$ Institut of Space and Astronautical Science, 3-1-1 Yoshinodai, Sagamihara, Kanagawa 22, Japan\\
$^3$ Institute of  Astronomy, Madingley Road, Cambridge CB3 0HA\\
$^4$ Department of Astronomy and Astrophysics, Penn State, 525 Davey Lab, University Park, PA 16802, USA\\
$^5$  Department of Astronomy, San Diego State University, San Diego, CA 92182, USA
}

\date{Accepted. Received 2003 May 30 }

\begin{document}
\label{firstpage}
\maketitle

\begin{abstract}
We report on the {\em XMM-Newton} timing properties of the most
X-ray variable, radio-quiet, Narrow-Line Seyfert 1 galaxy IRAS
13224--3809. 
IRAS 13224--3809 continues to display the extremely variable behavior that
was previously observed with {\em ROSAT} and {\em ASCA}; however, no giant, rapid
flaring events are observed.  We detect variations
by a factor as high as $\sim$8 during the 64 ks observation, and the 
variability is persistent throughout the light curve.  Dividing the light
curve into 9 minute segments we found almost all of the segments to be variable at
$>$ 3 $\sigma$.
When the time-averaged cross-correlation
function is calculated for the 0.3--0.8 keV band with the
3--10 keV band, the cross-correlation profile is skewed indicating a possible smearing of 
the signal to longer times (soft band leading the hard). 
A correlation between count rate and
hardness ratio is detected in four energy bands.  In three cases the
correlation is consistent with spectral hardening at lower count rates which can 
be explained in terms of a partial-covering model.
The other band displays the reverse effect, showing spectral hardening at 
higher count rates.  We can explain this trend as a more variable power-law component compared to
the soft component.
We also detect a delay between the 0.3--1.5~keV count rate and the  
0.8--1.5~keV to 0.3--0.8~keV hardness ratio, implying flux induced spectral variability. 
Such delays and asymmetries in the cross correlation functions could be suggesting
reprocessing of soft and hard photons. 
In general, much of the
timing behavior can be attributed to erratic eclipsing behavior associated with the partial covering 
phenomenon, in addition to intrinsic variability in the source.
The variability behavior of IRAS 13224--3809 suggests a 
complicated combination of effects which we have started to disentangle with this present 
analysis.

\end{abstract}


\section{Introduction}
The Narrow-Line Seyfert 1 galaxy (NLS1)
IRAS 13224--3809 is one of the most exciting NLS1s to study as it exhibits 
many of the extreme characteristics that make this class of objects so interesting.
Strong Fe~II emission in its optical spectrum (Boller et al. 1993); a giant soft X-ray
excess (Boller et al. 1996; Otani et al. 1996; Leighly 1999a; Boller et al. 2003);
and extreme, rapid, and persistent variability in X-rays 
(Boller et al. 1996; Boller et al. 1997; Otani et al. 1996; Leighly 1999b), are all
traits of IRAS 13224--3809.   

Boller et al. (2003; hereafter Paper~I) presented an array of new 
spectral complexities observed in an {\em XMM-Newton} observation.  In addition
to the strong soft excess component they detected a broad absorption feature at
$\sim$1.2 keV, first seen with {\em ASCA} (Otani et al. 1996; Leighly et al. 1997), and 
most probably due to Fe L absorption (Nicastro et al. 1999). 
Moreover, a sharp spectral power-law cut-off
at $\sim$8.1 keV was discovered, similar to the power-law cut-off at 7.1 keV observed in the NLS1 1H~0707-495
(Boller et al. 2002).

In this paper we have examined the timing properties of IRAS 13224--3809 as observed
with {\em XMM-Newton} and have related them to the spectral properties discussed in Paper~I.  
In the next section we will discuss the data analysis before
presenting the timing properties: OM light curve (Section 3), X-ray light curve and
flux variability (Section 4), spectral variability (Section 5), and a discussion of
our results in Section 6.
A value for the Hubble constant of $H_0$=$\rm 70\ km\ s^{-1}\ Mpc^{-1}$
and for the cosmological deceleration parameter of $q_0 = \rm \frac{1}{2}$ have
been adopted throughout.  Fluxes and luminosities were derived assuming the partial-covering
fit described in Paper~I and further assuming isotropic emission.

\section{The X-ray observation and data analysis}

IRAS 13224--3809 was observed with {\em XMM-Newton} (Jansen et al. 2001)
for $\sim$64 ks during revolution 0387 on 2002 January 19.  All instruments
were functioning normally during this time.
A detailed description of the observation and data analysis was presented
in Paper~I, and we will briefly summarize details which are important to the
analysis here.

The EPIC pn camera (Str\"uder et al. 2001) was operated in full-frame mode, and 
the two MOS cameras (MOS1 and MOS2; Turner et al. 2001) were operated in large-window
mode.  All of the EPIC cameras used the medium filter.
The Observation Data Files (ODFs) were processed to produce calibrated event
lists in the usual manner using the {\em XMM-Newton} Science Analysis System
(SAS) v5.3.
Light curves were extracted from
these event lists to search for periods of high background
flaring. A significant background flare was detected in the EPIC
cameras approximately 20 ks into the observation and lasting for
$\sim$5 ks. Although the total counts at flare maximum were at least a factor of 3 lower than 
the source counts, the segment was excluded during most of the analysis.  However,
when calculating cross correlation functions, we kept all of the data to avoid
dealing with gapped time series.
The total amount of good exposure time selected from the pn camera was 
55898 s.  

\section{The UV Observation}
The Optical Monitor (OM; Mason et al. 2001) collected data in the fast mode through 
the UVW2 filter (1800--2250\AA) for about the first 25 ks of the observation.  For the
remainder of the observation the OM was operated in the grism mode.
In total, seven photometric images were taken, each exposure lasting 2000 s.

The apparent UVW2 magnitude is 15.21 $\pm$ 0.07 corresponding to a band flux of
4.56 $\times$ 10$^{-15}$ erg s$^{-1}$ cm$^{-2}$ \AA$^{-1}$.  The flux is
consistent with previous UV observations with {\em IUE} 
(Mas-Hesse et al. 1994; Rodr\a'\i guez et al. 1997).
The RMS scatter in the seven hour OM light curve is less than 0.5\%.  
This agrees with the limits on rapid variability from the optical band
(Young et al. 1999; but see Miller et al 2000).

The optical to X-ray spectral index, $\alpha_{ox}$, was calculated using the definition
$$
\alpha_{ox} = \frac{log(f_{2 keV}/f_{1990 \AA})}{log(\nu_{2 keV}/\nu_{1990 \AA})},
$$
where $f_{2 keV}$ and $f_{1990 \AA}$ are the intrinsic flux densities at 2~keV and
1990~\AA, respectively.\footnote{A conversion to the standard definition of the spectral index 
between 2500 \AA~and 2 keV ($\alpha^{\prime}{_{ox}}$) is
$\alpha^{\prime}{_{ox}}$ = 0.96$\alpha_{ox}$ + 0.04$\alpha_{u}$, where $\alpha_{u}$ is
the power-law slope between 1990--2500 \AA.  We assume that $\alpha_{u}$$\approx$0 is
reasonable,
given the flatness of the UV spectra between 1100--1950 \AA (Mas-Hesse et al. 1994).} 
The selection of the UV wavelength corresponds to the  
peak response in the UVW2 filter (212 nm).  
Under the assumption that the UV light curve remains constant during the second
half of the observation, we determined an average power-law slope  
$<\alpha_{ox}>$ = --1.57.  As the 2 keV X-ray flux varies, $\alpha_{ox}$
fluctuates from --1.47 in the high state, to --1.73 in the low state.  

\section{The X-ray Light Curves}
In the remainder of the paper we will concentrate on the EPIC pn light curve due
to its high signal-to-noise.  
We will make reference to the various flux states as they
were defined in Paper~I: high ($>$2.5 counts s$^{-1}$), low ($<$1.5 counts s$^{-1}$), and
medium (1.5--2.5 counts s$^{-1}$).
The MOS light curve was also analyzed and found 
to be entirely consistent with the pn results.  In the interest of brevity, the MOS results will not
be presented here.

\subsection{Flux Variability}

\begin{figure}
       \psfig{figure=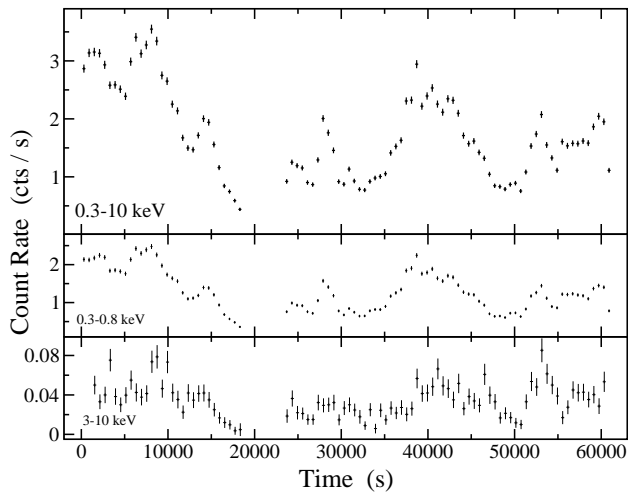,angle=-90,width=0.5\textwidth,clip=0}
      \caption{Three light curves of IRAS 13224--3809 in 600 s bins.
The energy bands (and average count rates) are (from top-down):
0.3--10 (1.74), 0.3--0.8 (1.30),  
and 3--10 (0.03) keV.  Zero seconds on the time axis
marks the start of the observation at 03:15:02 on 2002--01--19.  The region of 
high background flaring has been omitted. 
}
\label{xlc}
\end{figure}

EPIC pn light curves of IRAS 13224--3809 in three energy bands 
are presented in Figure~\ref{xlc}. 
The average count rate in the 0.3--10 keV energy band is (1.74 $\pm$ 0.05) counts
 s$^{-1}$.  Variations by a factor of $\sim$8 occur during the observation.
Using a 600 s bin size, the minimum and maximum 0.3--10 keV count rates are
(0.44 $\pm$ 0.03) and (3.55 $\pm$ 0.08) 
counts s$^{-1}$, respectively.  True to its strong soft-excess nature, IRAS 13224--3809
generates nearly three-quarters of the average count rate in the 0.3--0.8 keV
band.
Light curves were compared to a constant fit with a $\chi^2$ test.  A
constant fit could be rejected for all of the light curves at $>$3 $\sigma$ significance.

Giant-amplitude flaring events, as observed in the {\em ROSAT} and {\em ASCA}
observations, are not present during this observation.  In fact, the light curve
appears to be in a period of relative {\em quiescence} compared to those earlier
observations.  
The unabsorbed 0.3--2.4 keV flux is between (0.3--0.7) $\times$ 10$^{-11}$
erg s$^{-1}$ cm$^{-2}$ during the {\em average} low and high flux states.  Extrapolation of the
data from 0.3 keV to 0.1 keV allows us to estimate a 0.1--2.4 keV flux of (0.4--1.1) $\times$ 10$^{-11}$
erg s$^{-1}$ cm$^{-2}$.  It is interesting to note that while the {\em XMM-Newton} light curve does not display
giant flaring episodes, the flux during this observation   
is close to the $ROSAT$ HRI (0.1--2.4 keV) flux during the largest flaring events
(3.3 $\times$ 10$^{-11}$ erg s$^{-1}$ cm$^{-2}$; Boller et al. 1997). Therefore, in comparison to the
$ROSAT$ observations, IRAS 13224--3809 is in a relatively high flux state.  
In fact, the median HRI flux over the $ROSAT$ 30 day monitoring campaign was $\sim$ 0.2 $\times$ 10$^{-11}$
erg s$^{-1}$ cm$^{-2}$, while the average estimated 0.1--2.4 keV flux during this {\em XMM-Newton}
observation is about a factor of three greater. 
It is not clear,
with only the current observation in hand, whether this flux increase is a long-term change (i.e.
lasting for periods of days-months), or if we are observing the AGN during some sort of
extended outburst.  

\subsubsection{Rapid Variability}
\begin{figure}
       \psfig{figure=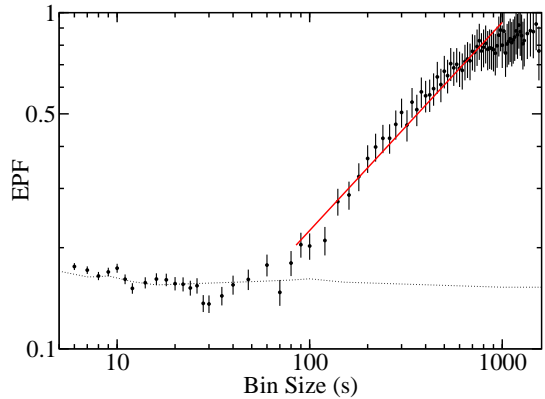,angle=-90,width=8.3cm,clip=}
      \caption{The EPF as a function of bin width calculated for IRAS 13224--3809.  The dotted line
is the estimated level of the EPF expected from Poisson noise.  The faint, solid line is
a power-law fit to the 85--1000s (1.2 $\times$ 10$^{-2}$--10$^{-3}$ Hz) region, with a normalisation
of 0.013$^{+0.003}_{-0.002}$ and an index of 0.619 $\pm$ 0.036.  This measured index is
related to the power-law index of the power spectrum ($\alpha$), and consistent with
$\alpha$ = -1.619 (see Yaqoob et al. 1997 for a detailed explanation).
}
\label{epf}
\end{figure}

We conducted a search for the shortest time scales for which we could detect variability
in IRAS 13224--3809 by employing the excess pair fraction method (Yaqoob et al. 1997).
We constructed 90 light curves with time bin sizes from 6--1550 s and calculated the 
excess pair fraction (EPF)
for each of them.  The results are plotted in Figure~\ref{epf}.  The dotted line in 
Figure~\ref{epf} is the level of the EPF expected from Poisson noise.  
The measurements run into the Poisson noise for bin sizes less than $\sim$70 s.
With a bin size of 90 s the source is variable at greater than 3 $\sigma$.

We searched further for indications of rapid variability by determining the shortest 
doubling times; that is, by searching for the shortest time interval in which the count rate doubles.
Count rate doubling times on the order of
800 s were reported during the $ROSAT$ observations and as low as 400 s during the $ASCA$ observation.  
With this most recent {\em XMM-Newton} observation, numerous such events are found on time scales 
between 300--800 s, and the shortest measured doubling time is about 270 s.

\subsubsection{Persistent Variability}
Clearly absent from the light curves are the giant and rapid flux outbursts that 
one has come to expect from IRAS 13224--3809
(Boller et al. 1993; Otani et al. 1996; Boller et al. 1997; Leighly 1999b;
Dewangan et al. 2002).
In the galaxy's defence, long monitoring with {\em ROSAT}
(Boller et al 1997) suggests that such flaring events occur every few days;
hence, were probably missed by this short ($<$ 1 day) observation.
However, short time binning of the light curve ($<$100 s) reveals  
persistent variability.  

With 60 s binning of the data, we searched for segments of quiescence in the
light curve.  Dividing the light curve into 10 minute segments, we used the $\chi^2$ 
test to search for global variability in each segment and found all of the segments to be
variable at $>$3 $\sigma$ significance.  In this manner, we worked
our way through shorter time segments.  Examining 9 minute segments we found that 101 of
103 segments were variable at $>$3 $\sigma$ significance (the other 2 segments
were variable at $>$2.6 $\sigma$).  The shortest segment size tested was 5 minutes,
for which we determined that $\sim$62\% (115/187) of the segments showed 
variability at a significance level $>$2.6 $\sigma$.
The variable segments do not appear to be associated with a particular
flux state. About 65\% of the high flux segments and 57\% of the low flux segments
displayed variability. 

The persistent variability on such short time scales 
disfavors the idea that the variability
is dominated by intrinsic disc instabilities, primarily because of the much longer time scales
expected for this process (Boller et al. 1997).  
This is, perhaps, inline with 
the lack of simultaneous UV fluctuations (as found in the OM light curve).
Some AGN models predict that the UV and soft X-ray emission are the same physical process,
namely thermal emission from the accretion disc ({\em Big Blue Bump}).  Bearing this in mind,
one would expect that the UV and soft X-rays should vary on similar time scales if the
variability were due primarily to disc instabilities.  Since this is not the case for IRAS 13224--3809
(in addition to the high thermal temperature derived in Paper~I), there must be some other
overriding process responsible for the rapid X-ray variability on time scales of minutes.

The persistent variability on such short time scales is also in contradiction with the idea
that the variability arises from a {\em single}, rotating hot spot on the disc. 
Furthermore, 
there are no detectable temperature changes in the soft spectrum, as would be 
expected due to beaming effects.
However, the hot spot model cannot be dismissed as the cause of
the large-amplitude fluctuations during flaring episodes.  
The persistent variability on time scales of minutes is most likely due to a combination of 
effects, and no one origin (e.g. disc instabilities, coronal flaring, hot spots) can be
definitively ruled out.

\subsubsection{Radiative Efficiency}
If we assume photon diffusion through a spherical mass of accreting matter we
can estimate the radiative efficiency, $\eta$, from the expression $\eta$ $>$ 
4.8 $\times$ 10$^{-43} (\Delta L/\Delta t)$ (Fabian 1979; see also Brandt et al.
1999 for a discussion of important caveats), where $\Delta L$ is the change in 
luminosity, and $\Delta t$ is the time interval 
in the rest-frame of the source.  From both {\em ROSAT} (Boller et al. 1997) and 
{\em ASCA} (Otani et al. 
1996) observations of IRAS 13224--3809, a radiative efficiency exceeding the
maximum Schwarzschild black hole radiative efficiency ($\eta$$=$0.057) was
determined. It is tempting to attribute this result to accretion onto a Kerr
black hole, or anisotropic emission, or X-ray hot spots on the disc.

We calculated the radiative efficiency at several points in the 200 s binned 
0.3--10 keV light curve.  From the mean unabsorbed luminosity (6 $\times$ 10$^{43}$ 
erg s$^{-1}$) and average count rate (1.74 $\pm$ 0.05 counts s$^{-1}$)
we were able to calculate a conversion factor between count rate and luminosity 
(1 count s$^{-1}$ = 3.4 $\times$ 10$^{43}$ erg s$^{-1}$) and
hence, we are able to calculate the luminosity rate of change from the observed change
in count rate.  Minimum and maximum count rates and times were averaged over at
least two bins (ideally three).
The regions selected correspond to the most rapid changes in count rate
(either rising or falling), and they occupy various regions in the 
light curve (i.e. regions were selected in the high, low, and medium flux states, 
as well as events transcending the different flux 
states).

The most rapid rate of change was $(\Delta L/\Delta t)$ = (4.10 $\pm$ 0.91) 
$\times$ 
10$^{40}$ erg s$^{-2}$ corresponding to a radiative efficiency of $\eta$ $>$ 0.020.
Averaging over all ten selected regions we determine an average radiative
efficiency of $\eta$ $>$ 0.013.  No clear trend is seen between
the value of $\eta$ and the flux state of the object from which the measurement
was made.  Since the calculated $\eta$ is a lower limit,
our small value is not inconsistent with the $\eta$ calculated from the {\em ROSAT}
or {\em ASCA} observations; nevertheless, it is nearly a factor of 10 smaller, and
we are clearly consistent with the efficiency
regime of a Schwarzschild black hole.  
It would stand to reason that we do not necessarily have
a Kerr black hole in IRAS 13224--3809 since radiative efficiency
limits in excess of the Schwarzschild limit are only found in the {\em ROSAT} (Boller et al. 1997)
and
{\em ASCA} (Otani et al. 1996) observations during periods of giant flaring outbursts. 
The large values of $\eta$ found in past
observations are probably due to radiative boosting or anisotropic emission during
the flaring events.

\subsubsection{Correlations among the light curves}
Several light curves in different energy bands were produced and compared with each other, and found to be correlated.
The significance of
the correlation was measured with the Spearman rank correlation coefficient
(Press et al. 1992; Wall 1996).  
All of the light curves are well correlated with each other at $>$99.9\%
confidence.

\begin{figure}
\begin{minipage}{8.3cm}
      \psfig{figure=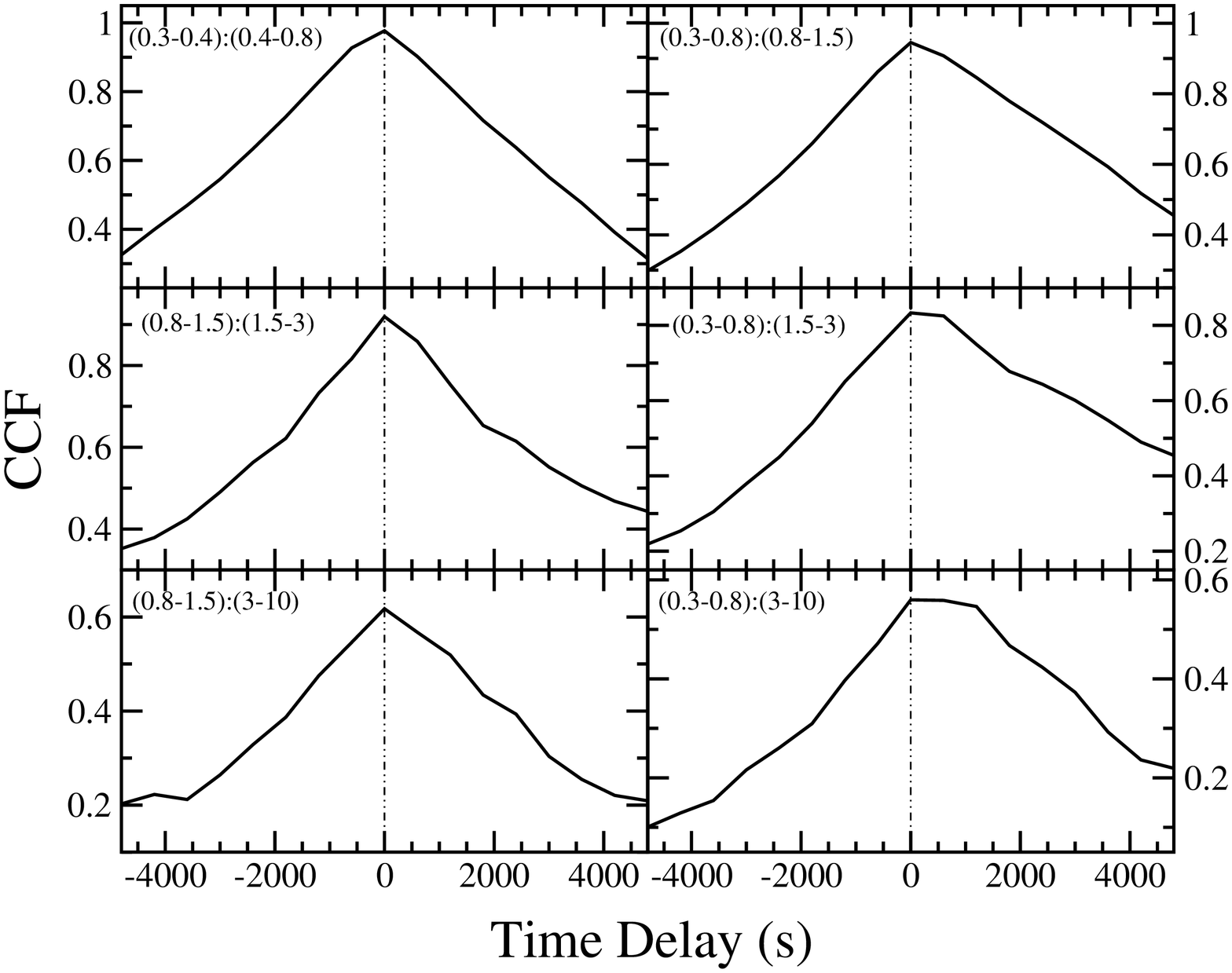,angle=-90,width=9.3cm,height=9.3cm,clip=}
      \caption{Six of the CCFs calculated from the light curves with 600s binning.
The energy bands being cross correlated are shown in the top left of each window. 
The calculations were conducted such that a positive shift in time would indicate
a lag of the second energy band.
}
\label{ccf_lc}
\end{minipage}
\begin{minipage}{8.3cm}
      \psfig{figure=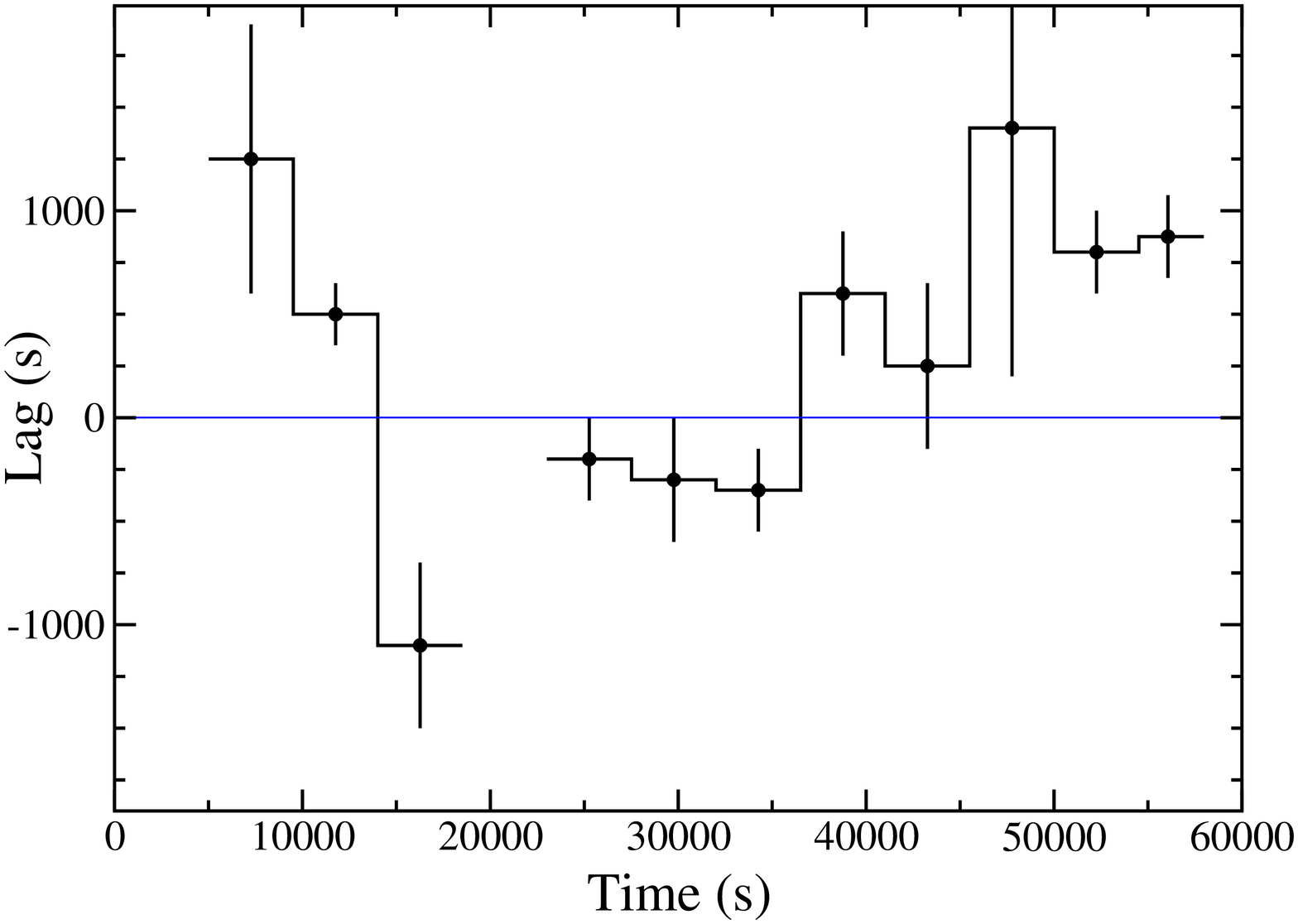,angle=-90,width=8.3cm,clip=}
      \caption{The lag measured between the 3--10 keV and 0.3--0.8 keV light curves, in 4500 s intervals,
over the duration of
the observation.  The lag alternates in such a way that at times the soft band 
leads the hard (positive lag), and at other times, the hard leads the soft (negative lag).  The time interval during which
the background flare occurs has been omitted.
}
\label{ccf_bill}
\end{minipage}
\end{figure}

Given the variability in all the light curves, and the significant correlation
amongst all of them, it was natural to search for leads or lags by calculating
the cross correlation function (CCF). We calculated CCFs between all the light
curves and we present six of them in Figure~\ref{ccf_lc}.
Most of the cross correlations are symmetric with a peak corresponding to zero
time delay.  
However, when the soft 0.3--0.8~keV band is cross correlated with harder
energy bands (right column of Figure~\ref{ccf_lc}), the CCFs 
become broad and somewhat
flat-topped with a noticeable asymmetry toward longer lags.
Because the time sampling is sufficiently fast to resolve the
variations in the light curves, it is the width of the ACF that ultimately
limits the resolution of the CCF. The soft light curve has a red
power spectrum (a power-law fit to the power spectrum has an index
of $\sim$~-2; see Figure~\ref{epf}), giving an ACF that is quite broad.
The pronounced asymmetry of the CCF suggests we are barely resolving a lag
between the two light curves.  Using the {\em centroid} of the CCF     
yields a lag of the 3--10 keV band by 460 $\pm$ 175 s.
The other two CCFs in the right column of Figure~\ref{ccf_lc} also reveal a slight
asymmetry.  The CCFs peak at a lag of zero, but the asymmetry suggests the hard bands lag 
the soft. The pronounced asymmetry of the 3--10 keV versus the 0.3--0.8 keV CCF warranted 
more attention. 

Careful inspection of the light curves indicate that the 
hard X-rays do not always lag the soft; in some cases it appears that
the reverse is true. To investigate further, we cut the soft and hard light curves into
12 segments, each 4500~s in duration with 50\% overlap between them
(so only ever other segment is independent). We then computed the CCF
in each of these shorter light curves. The results confirmed our visual
suspicion --- in some cases the hard X-rays lead the soft while at other  
times it lags. The results are shown in Figure~\ref{ccf_bill}.
The lags span approximately -1100~s to +1400~s and appear to be mildly
correlated with the soft X-rays in the sense that when the soft X-rays
are strong, the hard lags the soft. The lag is positive more often than
negative in our light curves, explaining why the full CCFs show a slight
positive lag.  
Error bars on the lags were determined via a Monte Carlo technique:
5000 simulated soft and hard light curves were generated by varying every
observed datum according to a Gaussian distribution whose standard
deviation was equal to the uncertainty in the observation.
The CCF was computed for each light curve pair and the peak recorded.
The width of the distribution of the 5000 peaks was used to estimate the 
uncertainty in the lag.  This was repeated for each of the 12 light curve
segments.
The perceived lags and leads are likely due to a physical separation between
the soft and hard emitting regions and/or reprocessing.

\section{Detection of significant and rapid Spectral Variability}
\begin{figure}
       \psfig{figure=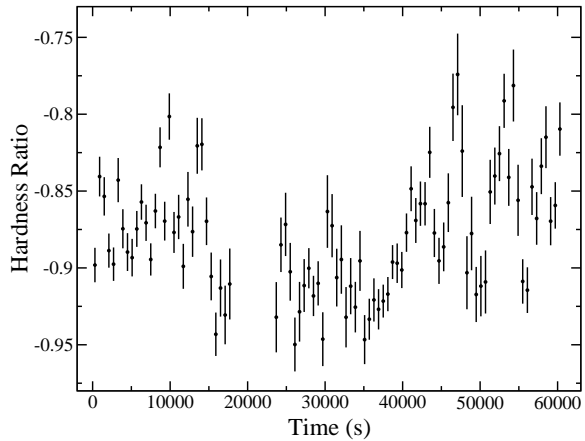,angle=-90,width=8.3cm,clip=}
      \caption{A typical hardness ratio variability curve. In this example $H$=1.5--10 keV and
$S$=0.3--1.5 keV (see text for details).  Zero seconds on the time axis
marks the start of the observation at 03:15:02 on 2002--01--19.  The region of
high background flaring has been omitted.
}
\label{hr}
\end{figure}

From the available light curves we calculated seven hardness ratios by the
definition $(H-S)/(H+S)$, where $H$ and $S$ are the count rates in the hard and
soft bands, respectively.  With this formalism, the value of the hardness ratio
will be between $-1$ and +1, and the harder spectra will have more positive values.
A typical hardness ratio is shown versus time in Figure~\ref{hr}.  In this case $H$=1.5--10 keV
and $S$= 0.3--1.5~keV.  All seven ratios show similar time variability to the hardness
ratio in Figure~\ref{hr}, albeit, some with poorer signal-to-noise.
In addition, each hardness ratio curve was tested for global variability via a $\chi$$^2$ test.
In all cases, the hardness ratios were inconsistent with a constant at greater than
2.6 $\sigma$.
The spectral variability depicted in Figure~\ref{hr} is significant and rapid.  In 
addition, the variability appears to intensify after about 40 ks into the observation.
With the multiple spectral components required in the spectral model, isolating the location
of the spectral variability is challenging.

\begin{figure}
       \psfig{figure=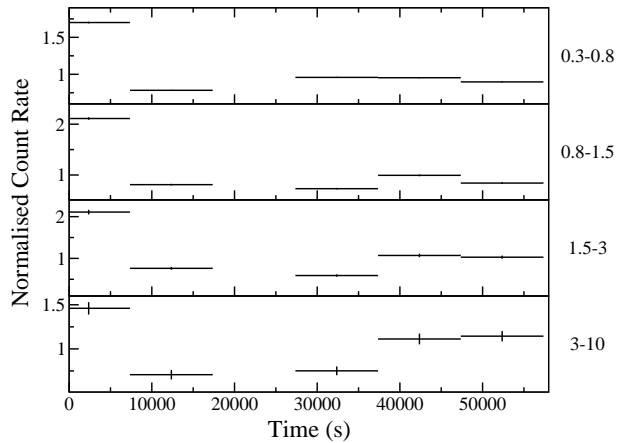,angle=-90,width=8.3cm,clip=}
      \caption{Four normalised X-ray light curves in various energy bands, and in 10 ks bins.
After $\sim$40 ks, the amplitude of the fluctuations becomes greater in the 3--10 keV band
then in the softer energy bands.  Note that although the count rate error bars are
plotted, they are too small to see in some of the light curves.
}
\label{lc10ks}
\end{figure}

We notice in Figure~\ref{xlc} that the overall trend in the count rate light curves is
very similar between different energy bands; however, on closer inspection it becomes
noticeable that the amplitude of the variations is different.
To illustrate this observation better we produce normalised light curves in
four energy bands, and in 10 ks bins (Figure~\ref{lc10ks}).  It becomes clear with this manner
of binning that the amplitude of the flux variations are different among the energy bands, and
indeed, there does appear to be a significant increase in the 3--10~keV flux after $\sim$40 ks.

\begin{figure}
       \psfig{figure=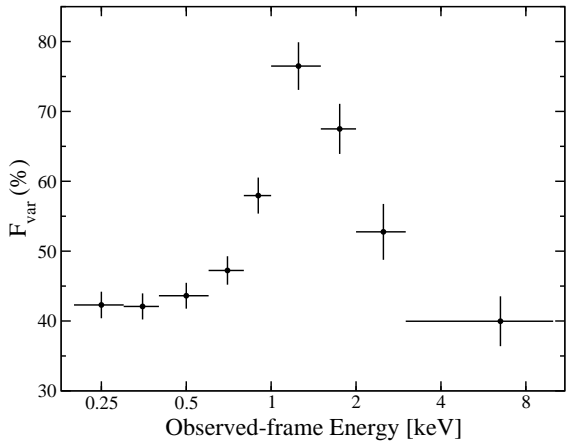,angle=-90,width=8.3cm,clip=}
      \caption{F$_{var}$ calculated in nine energy bins between 0.2 and 10 keV.
The light curves used were in bins of 200s.
}
\label{fvar}
\end{figure}

We searched for the source of the spectral variability further by calculating the fractional
variability amplitude (F$_{var}$) following Edelson et al. (2002).  
The results are shown in Figure~\ref{fvar} using 200 s binning of the light
curves.
Most of the variability is observed between 0.8 and 2 keV.
This region is quite complex as it includes the broad Fe L
absorption feature, and the point where the blackbody and power-law components
intersect (see Paper~I); hence, it is difficult to isolate any one component.  

A difference spectrum (high-low) also confirms the spectral variability.  A power-law is a good fit
to the difference spectrum over the 1.3--8 keV band.

\subsection{Flux correlated spectral variability}
\begin{figure}
       \psfig{figure=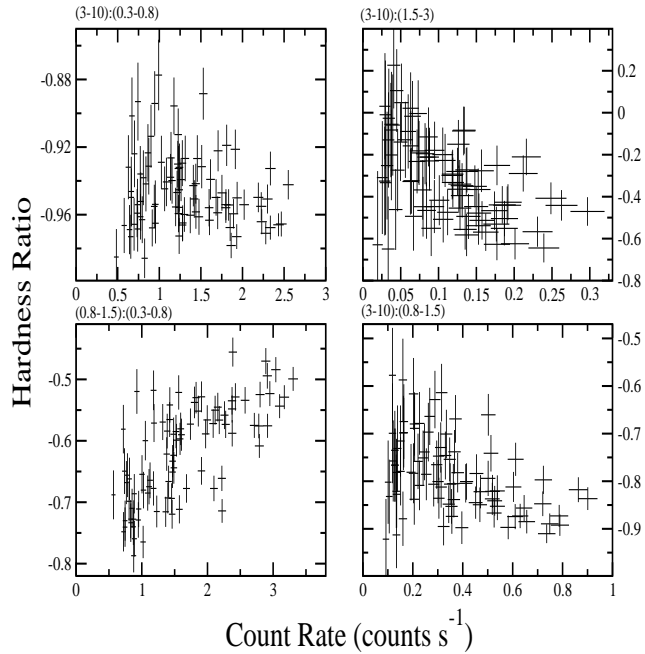,angle=-90,width=9.1cm,height=9.1cm,clip=}
      \caption{Four hardness ratios correlated with count rates.  The bands used
are show in the upper left of each panel.  The count rates used are the summation of $H$ 
and $S$.
}
\label{hrcr}
\end{figure}

Given the reasonable agreement in the appearance of the hardness ratios and the 
light curves we searched for common trends in hardness ratio versus count rate plots.
Using the Spearman rank correlation coefficient to test for correlations,
four of the hardness ratios were found to be correlated with the light curves at more than 99.5\%
confidence.  These four hardness ratios are plotted against the count rate in
Figure~\ref{hrcr}.
The four flux correlated hardness ratios are ($H$:$S$): \\
(1)  (3--10):(0.3--0.8),\\
(2)  (3--10):(1.5--3),\\
(3)  (0.8--1.5):(0.3--0.8),\\
(4)  (3--10):(0.8--1.5).\\
The count rates used are the summation of $H$ and $S$.

As can be seen from Figure~\ref{hrcr}, in four cases the spectral variability is correlated with the
variability in the count rate.  In three cases, spectral hardening was observed during 
periods of lower count rates.
This general effect has been observed previously in IRAS 13224--3809 (Dewangan et al. 2002), 
as well as in other AGN, for example: MCG--6--30--15 (Lee et al. 2000),
NGC 5548 (Chiang et al. 2000), 3C273 (Turner et al. 1990), and 
NGC 7314 (Turner 1987).  The trend is a more common feature among X-ray novae (see Tanaka \& Shibazaki
1996).  Such a trend is predicted by the partial-covering model, and is further supported by
the spectral analysis of IRAS 13224--3809 in Paper~I (see their figure 7).  
However, the hardness ratio (0.8--1.5):(0.3--0.8) (lower left panel of 
Figure~\ref{hrcr}) shows the reverse effect -- increased intensity  
when the spectrum is hard.  Such a relation was also found in the radio-loud NLS1
PKS 0558--504 (Gliozzi et al. 2001).  In that object, the hardening with increased flux
was interpreted as an additional hard component due to a radio jet.  This is probably
not the case for IRAS 13224--3809 since it is radio-quiet and since the unusual
hardness pattern is only observed over a small energy range.
Since the 0.8--1.5 keV energy band is very complicated due to the comparable contributions
from the thermal, power-law, and Fe L components, it is difficult to isolate or 
eliminate any one of these components.  
An alternative explanation for the observed trend is that each spectral component varies
differently (or not at all) with changing flux.
For example, the spectral variability in the Fe L could be independent of the flux.  
Similarly, the continuum in this band could also be independent of the flux changes and of the Fe L.  
However, when these possibly independent behaviors are combined, the
trend presented in Figure~\ref{hrcr} (lower left panel) arises.
A more likely picture is that the intrinsic power-law is
more variable than the soft component.  This is supported somewhat by Figure~\ref{lc10ks}, and
we provide further evidence in support of this claim in Table 1.  We modelled the 0.2--7~keV
spectrum in various flux states (flux states are defined in Section 4 and in Paper~I). 
In Table 1 we
show the unabsorbed fluxes of the power-law and blackbody components in the 0.8--1.5 keV band.
It is clear that the power-law does in fact show larger amplitude variations in this energy band.
While the difference between the blackbody flux in the low and high state is $\sim$2.5, the power-law
flux changes by more than a factor of 7.  This is most likely the cause for the trend observed
in the lower left panel of Figure~\ref{hrcr}.
\begin{table}
\caption[]{
Comparison of the blackbody and power-law unabsorbed fluxes in the 0.8--1.5 keV band.
Fluxes are in units of 10$^{-13}$ erg s$^{-1}$ cm$^{-2}$.  The associated uncertainty
in the fluxes is derived from the uncertainty in the count rates, and it is shown in brackets.  The fourth column is the ratio
of power-law flux over blackbody flux (column 3/column 2).
}
\label{tab1}
\begin{flushleft}
\begin{tabular}{l c c c }\hline\hline
Flux State & Blackbody Flux & Power-law Flux & Ratio \\ \hline
High & 7.97 (0.18) & 5.00 (0.11) & 0.63      \\
Medium & 4.89 (0.14) & 2.41 (0.07) & 0.49 \\ 
Low & 2.95 (0.20) & 0.67 (0.05) & 0.23  \\   \hline\hline
\end{tabular}
\end{flushleft}
\end{table}

\subsection{A possible lag between flux variations and spectral variations}
\begin{figure}
       \psfig{figure=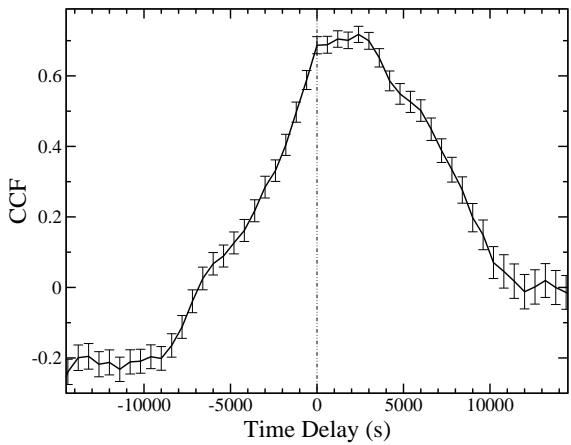,angle=-90,width=8.3cm,clip=}
      \caption{The cross correlation of the $H$=0.8--1.5 keV and $S$=0.3--0.8 keV light 
curve and hardness ratio variability curve.  The CCF indicates a short delay between the two curves, such
that, the hardness ratio curve lags behind the light curve. 
}
\label{hrlcccf}
\end{figure}

Further complexities arise when we scrutinize the 0.3--1.5 keV light curve and spectral variations 
in greater detail.  
We calculated the cross correlation function of the 0.3--1.5 keV light curve versus the
0.8--1.5 keV to 0.3--0.8 keV hardness ratio variability curve.  The cross correlation is presented in Figure~\ref{hrlcccf},
and indicates that the hardness ratio curve lags behind the light curve by 2000$^{+1500}_{-2000}$ s.  It is not clear
whether this observation is supporting the reprocessing of soft photons idea, or indicating variability
in the Fe L absorption feature, or some other effect.

Calculating light curve versus hardness ratio cross correlation functions for the other bands
did not reveal any significant correlations.  This could be an effect of the poorer 
signal-to-noise
in those bands as opposed to some physical process.  The band for which we have
determined a lag is, perhaps not surprisingly, the highest signal-to-noise energy range.

\section{Discussion}
\subsection{General Findings}
We have presented the timing properties of the NLS1 IRAS 13224--3809 
from a 64 ks {\em XMM-Newton} observation.
Our findings are summarized below.
\begin{itemize}
\item[(1)]
As expected, persistent and rapid variability was prevalent, although no giant-amplitude
variations are detected.  The source was in a relatively high flux state compared to the
{\em ROSAT} HRI observation.
\item[(2)]
Unlike previous {\em ROSAT} and {\em ASCA} observations, 
the radiative efficiency was below the limit for a Schwarzschild black hole.
\item[(3)]
Light curves in several bands were all well correlated.  When any two light curves were cross
correlated, most showed a symmetric correlation peaking at zero lag.  However,
when the time averaged 0.3--0.8~keV band was cross correlated with higher energy bands, 
an asymmetric cross-correlation profile was found.  Further inspection indicates that 
the lag varies; in some cases the hard X-rays lead, and at other times they lag.
\item[(4)]
Four of seven hardness ratio variability curves showed correlations with count rate
fluctuations.  Three of the four indicated spectral hardening at lower
count rates, supporting the partial-covering scenario.  The fourth hardness ratio, 
(0.8--1.5):(0.3--0.8), was characterized by hardening at higher count rates, most
probably due to a more variable power-law component.
\item[(5)]
The same hardness ratio, (0.8--1.5):(0.3--0.8), also shows a possible
time delay compared to the light curve; thus, suggesting flux induced spectral
variability.
\end{itemize}

\subsection{Reprocessing Scenario}
The basic idea in many AGN models is that high energy emission is produced in the
accretion disc corona by Comptonisation of soft photons, which are likely thermal in nature.  The detected
positive lags (soft leads hard) in Figure~\ref{ccf_lc} and~\ref{ccf_bill}, could be indicative of 
such a process.  In this situation the time lag would be due to either a physical separation
between the two emitting regions, or the reprocessing rate in the corona, or, most likely, a combination of the
two.  On the other hand, we also see evidence of the hard emission leading the soft (Figure~\ref{ccf_bill}
and~\ref{lc10ks}), which could be
occuring if high energy coronal photons are irradiating the disc and being Compton scattered.
In this 
case, the perceived lag would most probably be due to a physical separation between the emitting
regions, since the reprocessing rate is likely short due to the high densities in the disc.
Interesting is the fact that the apparent leads and lags are alternating, indicating that both processes
are occurring simultaneously but only one is dominating at a given time.

\subsection{Future tests for the Partial-Covering and Discline Models}
The partial-covering model (Holt et al. 1980; Boller et al. 2002; Tanaka et al 2003) is a 
very good fit to the spectrum in Paper~I.  The fit
can adequately describe the sharp spectral feature at 8.1 keV, and the absence of
the Fe line expected from the fluorescent yield, all with a reasonable Fe overabundance.  
The panels in Figure~\ref{hrcr} which demonstrate spectral hardening during states
of lower intensity are fully consistent with the partial-covering scenario.  
Observations of the intensity recovering above the edge, and the detection of 
other edges (e.g. Ni spectroscopy), would validate the partial-covering scenario in 
IRAS~13224--3809.

The discline model (Fabian et al. 2002) also provided a good fit across the whole energy
band to the
data in Paper~I, and without requiring extra absorption (albeit, with a large equivalent width 
for the broad Fe line).
Such a model is not inconsistent with the observed
flux and spectral variability in IRAS 13224--3809, nor is it inconsistent
with the interpretation that the asymmetric cross correlation functions are a result
of reprocessing of soft energy photons.  We could test for the presence of the
Fe line by measuring the depth of the edge feature at $\sim$ 8.1 keV in various flux or temporal
states.  The overwhelming strength of the proposed line suggests that the variations we have detected
probably arise within the line.  Therefore, one would expect that as the line profile changes
so would the depth of the edge.  A constant edge depth would be inconsistent with the
discline interpretation.  With the current photon statistics it is not possible to conduct
such a test precisely.

\vskip 0.2 in
In general, much of the timing behavior can be described in the context of the
partial covering phenomenon, with intrinsic source variability being a secondary effect.
The variability behavior of IRAS 13224--3809 suggests a complicated combination of effects
which we have started to disentangle in this present analysis.
A much longer observation with
{\em XMM-Newton}, including the detection of a giant outburst event, would allow us to answer 
several of the questions raised during this most recent X-ray observation.

\section*{Acknowledgements}
The authors are thankful to C.G. Dewangan for useful comments leading to the
improvement of this paper.
Based on observations obtained with XMM-Newton, an ESA science mission with
instruments and contributions directly funded by ESA Member States and
the USA (NASA).  WNB and ACF acknowledge support from NASA LTSA grant NAG5-13035
and the Royal Society, respectively.

\end{document}